\documentclass[conference, twocolumn]{IEEEtran}
\IEEEoverridecommandlockouts
\usepackage{setspace}
\usepackage{cite}
\usepackage{amsmath,amssymb,amsfonts}
\usepackage{algorithmic}
\usepackage{graphicx}
\usepackage{textcomp}
\usepackage{xcolor}
\usepackage{booktabs}
\usepackage{array}
\usepackage{siunitx}
\usepackage{multirow}
\usepackage{url}
\usepackage{tikz}
\usetikzlibrary{arrows.meta, positioning, shapes.geometric, calc, fit, backgrounds}

\def\BibTeX{{\rm B\kern-.05em{\sc i\kern-.025em b}\kern-.08em
    T\kern-.1667em\lower.7ex\hbox{E}\kern-.125emX}}

\usepackage{fancyhdr}
\usepackage{lipsum} 

\fancypagestyle{arxivcopyright}{%
  \fancyhf{} 
  \fancyfoot[C]{\footnotesize \textcopyright~2026 IEEE. Personal use of this material is permitted. Permission from IEEE must be obtained for all other uses, in any current or future media, including reprinting/republishing this material for advertising or promotional purposes, creating new collective works, for resale or redistribution to servers or lists, or reuse of any copyrighted component of this work in other works.}
}

\begin{document}

\title{On the Adversarial Vulnerability of Neural Receivers with Unrolled LDPC Decoders%
}

\author{%
\IEEEauthorblockN{Aakash Saini$^{\ast\ddagger}$, Jafar Aco Mohammadi$^{\ast}$, and Wolfgang Gerstacker$^{\ddagger}$}
\IEEEauthorblockA{$^{\ast}$\textit{Nokia, Munich, Germany} \\
$^{\ddagger}$ \textit{Institute for Digital Communications, Friedrich-Alexander-Universit\"at Erlangen-N\"urnberg, Erlangen, Germany}}

}

\maketitle

\thispagestyle{arxivcopyright} 

\begin{abstract}
Neural receivers for 5G New Radio (NR) match classical model-based receivers in link-level performance, but their robustness to adversarial perturbations at the receiver input remains underexplored. We show that standard gradient-based attacks applied end-to-end through an unrolled low-density parity-check (LDPC) decoder fail to significantly outperform a matched-power Gaussian perturbation. We identify the failure as gradient masking caused by $\mathrm{tanh}$ saturation in the boxplus operation of the belief-propagation (BP) check-node update, with a per-iteration gradient norm decay of more than four orders of magnitude across ten unrolled BP iterations. Based on this finding, we propose two structurally independent loss functions that restore attack efficacy. $\mathcal{L}_{\mathrm{BP}}$ routes the gradient through the unsaturated early iterations via per-iteration supervision, while $\mathcal{L}_{\mathrm{LLR}}$ is a margin loss on pre-decoder log-likelihood ratios (LLRs) whose gradient bypasses the LDPC decoder entirely. Both loss functions increase the block-error rate (BLER) by one to four orders of magnitude at matched perturbation power, with the results reproduced on both the Sionna NeuralReceiver and DeepRx. At low PSR, $\mathcal{L}_{\mathrm{LLR}}$-optimized perturbations achieve up to three orders of magnitude higher BLER on unseen channel realizations than $\mathcal{L}_{\mathrm{BP}}$-optimized perturbations.
\end{abstract}

\begin{IEEEkeywords}
adversarial attacks, NeuralReceiver, DeepRx, LDPC, belief propagation, obfuscated gradients, 5G NR.
\end{IEEEkeywords}

\section{Introduction}
Neural-network-based receivers (NNRs) have emerged as strong candidates for physical-layer signal processing in 5G New Radio (NR) due to their ability to jointly learn channel estimation, equalization and demapping. Architectures such as DeepRx~\cite{honkala2021deeprx} and the Sionna NeuralReceiver~\cite{hoydis2022sionna} achieve link-level performance comparable to classical model-based receivers. As NNRs advance toward deployment, their vulnerability to adversarial perturbations of the received signal emerges as an operational concern. An attacker injecting a small perturbation at the receiver antennas can exploit the non-linear structure of the learned receiver to cause significant decoding errors.

Adversarial vulnerability of deep-learning (DL)-based wireless receiver components has been studied in modulation classification~\cite{sadeghi2019adversarial}, semantic  communications~\cite{sagduyu2023vulnerabilities}, deep neural network (DNN)-based channel estimation for orthogonal frequency division multiplexing (OFDM)~\cite{bahramali2021robust}, and short-block neural decoders~\cite{shen2025verify,lei2026fragility}. Shen and Li~\cite{shen2025verify} applied interval analysis to verify the robustness of neural decoders on the AWGN channel with BCH codes, while Lei \emph{et al.}~\cite{lei2026fragility} have studied the fragility of transformer-based neural channel decoders under small AWGN channel perturbations. These works address standalone decoders over simplified wireless channel models, excluding NNRs integrated with unrolled LDPC belief propagation (BP) decoders. 

The fast gradient method (FGM), an $\ell_2$ adaptation of the fast gradient sign method (FGSM)~\cite{goodfellow2015explaining}, and projected gradient descent (PGD)~\cite{madry2018towards} are the standard single-step and multi-step adversarial attacks. The common assumption that computer-vision adversarial attacks transfer directly to wireless receivers fails for NNRs incorporating unrolled LDPC decoders in the back-propagation path. A standard end-to-end application of $\ell_2$-FGM or PGD (PGD-E2E) against an NNR with an unrolled 5G NR LDPC BP decoder yields negligible degradation in the BLER; moreover, a matched-power Gaussian jammer achieves similar performance to such gradient-based attacks at low or moderate PSR. We identify this as gradient masking~\cite{athalye2018obfuscated}, not genuine robustness of the NNR.

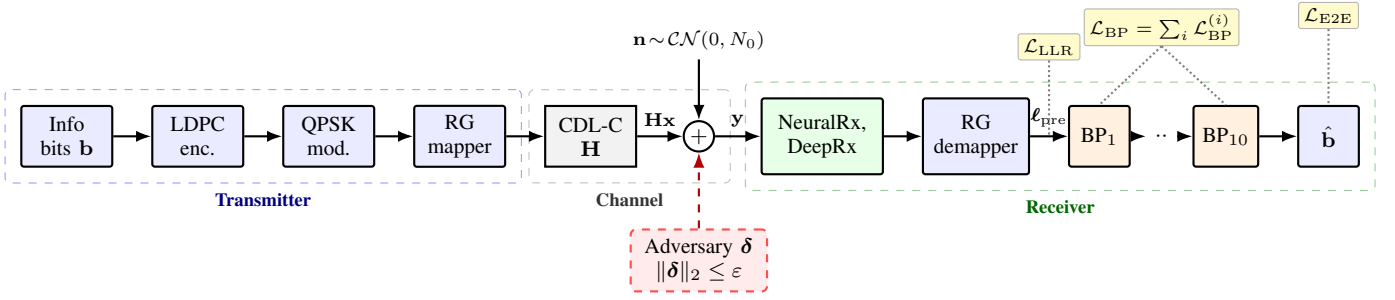
\begin{figure*}[!t]
\centering
\resizebox{\textwidth}{!}{%
\begin{tikzpicture}[
    >={Latex[length=2mm,width=1.5mm]},
    font=\footnotesize,
    node distance=4mm and 5mm,
    block/.style={
        draw, rectangle, rounded corners=1pt,
        minimum height=8mm, minimum width=12mm,
        align=center, thick, fill=blue!8
    },
    rxblock/.style={
        draw, rectangle, rounded corners=1pt,
        minimum height=10mm, minimum width=16mm,
        align=center, thick, fill=green!10
    },
    demapblock/.style={
        draw, rectangle, rounded corners=1pt,
        minimum height=10mm, minimum width=14mm,
        align=center, thick, fill=blue!8
    },
    bpblock/.style={
        draw, rectangle, rounded corners=1pt,
        minimum height=8mm, minimum width=8mm,
        align=center, thick, fill=orange!12
    },
    outblock/.style={
        draw, rectangle, rounded corners=1pt,
        minimum height=8mm, minimum width=8mm,
        align=center, thick, fill=blue!8
    },
    chan/.style={
        draw, rectangle,
        minimum height=8mm, minimum width=12mm,
        align=center, thick, fill=gray!10
    },
    sumop/.style={
        draw, circle, inner sep=0pt, minimum size=4mm, thick, fill=white
    },
    deltabox/.style={
        draw, dashed, thick, rectangle, rounded corners=2pt,
        minimum height=8mm, minimum width=18mm,
        align=center, fill=red!8, draw=red!70
    },
    losslabel/.style={
        draw=black!30, fill=yellow!25, rounded corners=1pt,
        inner sep=2pt, font=\scriptsize
    },
    arrow/.style={->, thick},
    dasharrow/.style={->, dashed, thick, red!70!black},
]

\node[block] (bits)    {Info\\bits $\mathbf{b}$};
\node[block, right=of bits] (enc)   {LDPC\\enc.};
\node[block, right=of enc]  (mod)   {QPSK\\mod.};
\node[block, right=of mod]  (rg)    {RG\\mapper};
\node[chan, right=of rg]    (chan)  {CDL-C\\$\mathbf{H}$};
\node[sumop, right=6mm of chan] (sum) {$+$};
\node[rxblock, right=6mm of sum] (nrx) {NeuralRx, \\DeepRx};
\node[demapblock, right=of nrx] (demap) {RG\\demapper};
\node[bpblock, right=of demap]  (bp1)  {BP$_1$};
\node[right=2mm of bp1, font=\footnotesize] (bpdots) {$\cdot \cdot$};
\node[bpblock, right=2mm of bpdots] (bpN) {BP$_{10}$};
\node[outblock, right=of bpN] (out) {$\hat{\mathbf{b}}$};

\draw[arrow] (bits) -- (enc);
\draw[arrow] (enc)  -- (mod);
\draw[arrow] (mod)  -- (rg);
\draw[arrow] (rg)   -- (chan);
\draw[arrow] (chan) -- node[above, font=\scriptsize] {$\mathbf{Hx}$} (sum);
\draw[arrow] (sum)  -- node[above, font=\scriptsize] {$\mathbf{y}$} (nrx);
\draw[arrow] (nrx)  -- (demap);
\draw[arrow] (demap)  -- node[above, font=\scriptsize] {$\boldsymbol{\ell}_{\mathrm{pre}}$} (bp1);
\draw[arrow] (bp1)  -- (bpdots);
\draw[arrow] (bpdots) -- (bpN);
\draw[arrow] (bpN)  -- (out);

\node[font=\scriptsize, above=8mm of sum] (noise) {$\mathbf{n}\!\sim\!\mathcal{CN}(0,N_0)$};
\draw[arrow] (noise) -- (sum);

\node[deltabox, below=10mm of sum] (delta) {Adversary $\boldsymbol{\delta}$\\$\|\boldsymbol{\delta}\|_2 \le \varepsilon$};
\draw[dasharrow] (delta) -- (sum);

\coordinate (lpre_pt) at ($(demap.east)!0.5!(bp1.west)$);
\node[losslabel, above=10mm of lpre_pt] (lllr) {$\mathcal{L}_{\mathrm{LLR}}$};
\draw[densely dotted, gray, thick] (lllr.south) -- (lpre_pt);

\node[losslabel, above=10mm of bpdots] (lbp) {$\mathcal{L}_{\mathrm{BP}}=\textstyle\sum_i\mathcal{L}_{\mathrm{BP}}^{(i)}$};
\draw[densely dotted, gray, thick] (lbp.south) -- (bp1.north);
\draw[densely dotted, gray, thick] (lbp.south) -- (bpN.north);

\node[losslabel, above=10mm of out] (le2e) {$\mathcal{L}_{\mathrm{E2E}}$};
\draw[densely dotted, gray, thick] (le2e.south) -- (out.north);

\begin{scope}[on background layer]
    \node[fit=(bits)(rg), draw=blue!40, dashed, rounded corners=2pt,
          inner sep=2mm,
          label={[blue!50!black, font=\scriptsize\bfseries]below:Transmitter}] {};
    \node[fit=(chan)(sum), draw=gray!50, dashed, rounded corners=2pt,
          inner sep=2mm,
          label={[gray!40!black, font=\scriptsize\bfseries]below:Channel}] {};
    \node[fit=(nrx)(demap)(bp1)(bpN)(out), draw=green!50!black!40, dashed, rounded corners=2pt,
          inner sep=2mm,
          label={[green!40!black, font=\scriptsize\bfseries]below:Receiver}] {};
\end{scope}

\end{tikzpicture}%
}
\caption{5G NR uplink system model with a neural network receiver (NNR), resource-grid demapping, and unrolled belief propagation (BP) decoding ($N_{\mathrm{BP}}=10$), where an adversary injects an $\ell_2$-bounded perturbation $\boldsymbol{\delta}$ at the receiver antennas. The NNR outputs resource-grid log-likelihood ratios (LLRs), demapped and serialized into the pre-decoder vector $\boldsymbol{\ell}_{\mathrm{pre}}$. Three gradient-based attack objectives are evaluated: end-to-end loss $\mathcal{L}_{\mathrm{E2E}}$ (post-decoder), iterative BP loss $\mathcal{L}_{\mathrm{BP}}$, and pre-decoder LLR loss $\mathcal{L}_{\mathrm{LLR}}$ (bypassing the BP graph).}
\label{fig:system}
\end{figure*}
PyJama~\cite{ulbricht2024pyjama} employs a Nachmani-style~\cite{nachmani2018deep} multi-iteration loss function within Sionna for differentiable jammer power-allocation optimization. Our work differs in purpose, formulation, and target system. PyJama optimizes a 14-dimensional OFDM power allocation via gradient descent to train a learnable jammer against classical MIMO receivers. We design a 3584-dimensional complex $\ell_2$-bounded perturbation against NNRs combined with unrolled LDPC decoders. The multi-iteration loss function in PyJama is deployed empirically as a tool, not as a countermeasure for gradient-masking. Our contribution is the gradient-masking diagnosis, quantification of the saturation effect, and the demonstration that a structurally unrelated second loss function recovers the attack efficacy.
The main contributions of this paper are as follows.
\begin{itemize}
    \item We demonstrate that standard $\ell_2$-FGM and $\ell_2$-PGD attacks optimized for the end-to-end binary cross-entropy (BCE) loss function with logits fail to significantly outperform a matched-power Gaussian jammer (baseline) against NNRs with unrolled 5G NR LDPC decoders.
    
    \item We provide the first identification of obfuscated gradients as the cause of the failure of end-to-end adversarial attacks against NNRs with unrolled LDPC BP decoders. While vanishing gradients in unrolled BP decoders have been previously observed during the training of neural decoders~\cite{dai2021learning}, we connect them to the obfuscated-gradients framework of Athalye \emph{et al.}~\cite{athalye2018obfuscated} for adversarial attacks.
    
    \item We propose two structurally independent bypass loss functions for PGD against unrolled BP decoders. $\mathcal{L}_{\mathrm{BP}}$ adapts the multi-iteration supervision of Nachmani \emph{et al.}~\cite{nachmani2018deep}, originally proposed for training, to the adversarial attacks. $\mathcal{L}_{\mathrm{LLR}}$ is a margin loss on pre-decoder LLRs that corrupts the BP decoder's inputs while its gradient bypasses the BP graph entirely. To our knowledge, this is the first attack loss function to exploit BP decoder sensitivity to pre-decoder LLR sign and confidence. Both loss functions increase the post-decoder BLER by up to four orders of magnitude over PGD-E2E.

    \item We verify architecture-agnostic gradient obfuscation and recovery. The same failure of the adversarial attacks based on end-to-end loss function and the same recovery under $\mathcal{L}_{\mathrm{BP}}$ and $\mathcal{L}_{\mathrm{LLR}}$ are observed on both DeepRx~\cite{honkala2021deeprx} and the Sionna NeuralReceiver~\cite{hoydis2022sionna}. We further observe that PGD attacks based on $\mathcal{L}_{\mathrm{LLR}}$ transfer across channel realizations with substantially higher BLER than attacks based on $\mathcal{L}_{\mathrm{BP}}$.
    
\end{itemize}

\section{System Model and Threat Model}\label{sec:system}
\subsection{System Model}
\label{sec:system_a}
We consider a 5G NR uplink transmission setting over a single-input multiple output (SIMO) CDL-C channel, modeled and simulated using the Sionna PHY~\cite{hoydis2023sionna} link-level simulator. The transmitter is equipped with a single antenna and the receiver with $N_r$ antennas. Let $\mathbf{b} \in \{0,1\}^k$ denote the information bit vector. It is encoded by a 5G NR low-density parity-check (LDPC) encoder of code rate $R = k/n$, producing the codeword $\mathbf{c} \in \{0,1\}^n$. The codeword is modulated using quadrature phase-shift keying (QPSK) and the resulting symbols are mapped onto an OFDM resource grid spanning $T$ OFDM symbols and $F$ subcarriers. Let $x_{t,f} \in \mathbb{C}$ denote the transmitted symbol on subcarrier $f$ of the OFDM symbol $t$, with $t=1, \ldots, T$, and $f=1, \ldots, F$. After propagation through the CDL-C channel, the $r$-th receive antenna observes 
\begin{equation}
y_{r,t,f} = h_{r,t,f} \cdot x_{t,f} + n_{r,t,f}+\delta_{r,t,f},
\label{eq:received}
\end{equation}
where $h_{r,t,f} \in \mathbb{C}$ denotes the wireless channel coefficient, $n_{r,t,f} \sim \mathcal{CN}(0, N_0)$ represents the AWGN at the $r$-th receive antenna, and $\delta_{r,t,f} \in \mathbb{C}$ stands for an adversarial perturbation defined in Sec.~\ref{sec:system_c}. The resource grid is the set of received signals $\{y_{r,t,f} \mid r = 1,\ldots,N_r,\ t = 1,\ldots,T,\ f = 1,\ldots,F\}$, indexed across $N_{\mathrm{c}} = N_r \cdot T \cdot F$ complex resource elements. 
We denote the perturbation vector formed by the vectorization of $\{\delta_{r,t,f}\}$ as $\boldsymbol{\delta} \in \mathbb{C}^{N_{\mathrm{c}}}$. The system operates at carrier frequency $f_c$ and at energy-per-bit to noise spectral density ratio $E_b/N_0$, where $E_b$ is the energy per information bit and $N_0$ is the noise spectral density. We operate at high SNR, where pre-decoder LLR magnitudes are large, placing the BP decoder in the severe $\tanh$ saturation regime, the worst case for gradient-based attacks. The complete end-to-end signal processing chain is illustrated in Fig.~\ref{fig:system}.
\subsection{Neural network-based Receivers} \label{sec:system_b}
We evaluate two NNRs of substantially different scale and architectural design. The first is the Sionna NeuralReceiver~\cite{hoydis2022sionna}, comprising a deep stack of convolutional layers with implicit channel estimation and approximately $4.86$\,M trainable parameters. The second is DeepRx~\cite{honkala2021deeprx}, a U-Net-based architecture that accepts an explicit least-squares channel estimate as an auxiliary input feature. We adopt a reduced-capacity variant comprising $7$ residual blocks, rather than the $11$ blocks of the original model, yielding approximately $0.64$\,M trainable parameters. Both architectures were trained with the AdamW optimizer and a one-cycle learning-rate scheduler for 30{,}000 iterations on randomly drawn information bit blocks at varying $E_b/N_0$.

The NNR produces the set of resource-grid LLRs $\{\ell^{\mathrm{RG}}_{q,t,f} \mid q = 1,\ldots,Q,\ t = 1,\ldots,T,\ f = 1,\ldots,F\}$, where $q$ indexes the bit position within a QPSK symbol. A resource-grid demapper extracts the coded-bit LLRs and serializes them into the pre-decoder LLR vector $\boldsymbol{\ell}_{\mathrm{pre}} \in \mathbb{R}^n$, with one entry per coded bit. The LLR vector is passed to the 5G NR LDPC decoder implemented in Sionna PHY, configured with the exact boxplus check-node update~\cite{hagenauer1996iterative}. 
We unroll the decoder across $N_{\mathrm{BP}}$ number of BP iterations to extract the per-iteration LLR vector $\boldsymbol{\ell}^{(i)}_{\mathrm{post}}$ for $i = 1, \ldots, N_{\mathrm{BP}}$. This unrolling provides the per-iteration gradient access required by the multi-iteration loss function introduced in Sec.~\ref{sec:attacks}. The post-decoder estimate of $\mathbf{b}$ is denoted by $\hat{\mathbf{b}}$.

\subsection{Threat Model and Adversarial Optimization}\label{sec:system_c}
We consider two threat models: the per-realization 
white-box and the transfer mode. In the per-realization white-box model, the adversary has full knowledge of the NNR weights and the channel realization $\{h_{r,t,f}\}$. In the transfer mode evaluated in Sec.~\ref{sec:results}, the adversary designs a perturbation on one set of channel realizations and applies it to unseen realizations. The per-realization model is the strongest available attacker; the transfer mode is a more practical setting. 
Given a chosen attack loss function $\mathcal{L}$, the per-realization adversary solves
\begin{equation}
\boldsymbol{\delta}^{\ast} = \arg\max_{\boldsymbol{\delta}} \, \mathcal{L}\bigl(\boldsymbol{\delta}\bigr) \quad \text{s.t.} \quad \|\boldsymbol{\delta}\|_2 \leq \varepsilon,
\label{eq:adversarial_opt}
\end{equation}
where $\mathcal{L}$ is evaluated on the receiver outputs computed from the perturbed signal $\{y_{r,t,f}\}$ in~\eqref{eq:received} via the NNR and unrolled BP decoder. The perturbation budget $\varepsilon$ is set by the perturbation-to-signal ratio (PSR) via
\begin{equation}
\varepsilon^2 = N_{\mathrm{c}} \cdot 10^{\mathrm{PSR}_{\mathrm{dB}}/10},
\label{eq:eps_from_psr}
\end{equation}
under the convention of unit average power per transmitted resource element. Noise is not included in the PSR definition, following the convention in~\cite{ulbricht2024pyjama,shen2025verify}.

The perturbation is injected directly at the receiver antennas, as in~\eqref{eq:received}, which is an abstraction of realistic jamming. In practice, a jammer transmits through its own jammer-to-receiver channel, which constrains the achievable perturbation directions and enables the receiver to estimate the jamming channel via pilot-based probing or channel reciprocity. We omit this coupling; our results therefore establish a vulnerability of the receiver model rather than a realizable over-the-air attack. We assume perfect timing and frequency synchronization between the adversary and the legitimate transmitter.

\section{Adversarial Attacks}\label{sec:attacks}
\subsection{Loss Functions}\label{sec:attacks_a}
We consider three loss functions for designing the adversarial perturbation. Each loss function intervenes at a distinct node in the NNR integrated with unrolled BP message-passing graph, enabling the mechanistic analysis presented in Sec.~\ref{sec:mechanism}. Before defining the loss functions, we fix notation for the receiver outputs.

\textit{Receiver outputs:} Given the perturbed received signal $\{y_{r,t,f}\}$ defined in~\eqref{eq:received}, the signal processing chain in Sec.~\ref{sec:system_b} produces three quantities used by the loss functions below: the pre-decoder LLR vector $\boldsymbol{\ell}_{\mathrm{pre}} \in \mathbb{R}^n$, the per-iteration post-decoder LLR vectors $\boldsymbol{\ell}_{\mathrm{post}}^{(i)} \in \mathbb{R}^n$ for $i = 1, \ldots, N_{\mathrm{BP}}$, and the information-bit LLR vector $\boldsymbol{\ell}_{\mathrm{info}} \in \mathbb{R}^k$ obtained by restricting the final-iteration LLRs to the $k$ systematic positions.

\textit{End-to-end loss function:} The end-to-end loss function applies BCE function with logits to the post-decoder information-bit LLRs,
\begin{equation}
\begin{aligned}
    \mathcal{L}_{\mathrm{E2E}} = -\frac{1}{k} \sum_{j=1}^{k} 
    \Bigl[ &b_j \log \sigma(\ell_{\mathrm{info},j}) +\\
    & (1-b_j) \log\bigl(1 - \sigma(\ell_{\mathrm{info},j})\bigr) \Bigr],
\end{aligned}
\label{eq:le2e}
\end{equation}
where, $b_j \in \{0,1\}$ is the $j$th true information bit and $\sigma(\cdot)$ is the element-wise sigmoid, and $\log$ denotes the natural logarithm.  
The gradient $\nabla_{\boldsymbol{\delta}} \mathcal{L}_{\mathrm{E2E}}$ is back-propagated through the information-bit extraction, all $N_{\mathrm{BP}}$ unrolled BP iterations, and the NNR, with respect to $\boldsymbol{\delta}$.

\textit{Multi-iteration BP loss function:} The second loss function sums per-iteration BCE-with-logits terms across all unrolled BP iterations, applied to the codeword LLRs at each BP iteration,
\begin{equation}\label{eq:lbp}
\begin{aligned}
\mathcal{L}_{\mathrm{BP}} &= \sum_{i=1}^{N_{\mathrm{BP}}} 
\mathcal{L}_{\mathrm{BP}}^{(i)}, \\
\mathcal{L}_{\mathrm{BP}}^{(i)} &= -\frac{1}{n} \sum_{j=1}^{n} 
\Bigl[ c_j \log \sigma(\ell_{\mathrm{post},j}^{(i)}) + \\
&\hspace{2cm} (1-c_j) \log\bigl(1 - \sigma(\ell_{\mathrm{post},j}^{(i)})\bigr) \Bigr],
\end{aligned}
\end{equation}
where, $c_j \in \{0,1\}$ is the $j$th true coded bit. The similar per-iteration supervision was introduced for training neural BP decoders by Nachmani \emph{et al.}~\cite{nachmani2018deep} to combat vanishing gradients across deep unrolled BP iterations. In this paper, we employ per-BP-iteration supervision within the adversarial-attack setting. Each per-BP-iteration term $\mathcal{L}_{\mathrm{BP}}^{(i)}$ measures the BP output at iteration $i$, evaluated on the full codeword. The gradients of early-BP-iteration terms are back-propagated through fewer BP steps and contribute substantially to $\nabla_{\boldsymbol{\delta}} \mathcal{L}_{\mathrm{BP}}$, while later-BP-iteration terms contribute negligibly to $\nabla_{\boldsymbol{\delta}} \mathcal{L}_{\mathrm{BP}}$.

\textit{Pre-decoder margin loss function:} The pre-decoder LLR vector is a natural target for adversarial perturbation. Under the Gaussian LLR consistency model of ten Brink~\cite{tenbrink2001convergence}, the LLR distribution at the decoder input characterizes the effective channel SNR, with post-decoder bit error rate expressible as a function of LLR mean and variance. 
The post-decoder error rate is therefore largely determined by 
the pre-decoder LLR distribution. Let $s_j = 2 c_j - 1 \in \{-1, +1\}$ encode the sign of the $j$-th true coded bit. Each coded bit has the correct sign at the decoder input when $s_j \cdot \ell_{\mathrm{pre},j} > 0$. We define a loss function that activates whenever the receiver fails to assign the correct sign to a bit with confidence at least $m$,
\begin{equation}
\mathcal{L}_{\mathrm{LLR}} = \frac{1}{n} \sum_{j=1}^{n} 
\max\!\left\{0,\, m - s_j \cdot \ell_{\mathrm{pre},j}\right\}.
\label{eq:lllr}
\end{equation}
The hinge is active whenever $s_j \cdot \ell_{\mathrm{pre},j} < m$ and exactly zero otherwise. The margin $m$ controls selectivity: small $m$ focuses the attack on bits near the decision boundary, while large $m$ broadens it to include weakly-correct bits. A structurally analogous hinge appears in adversarial classification~\cite{carlini2017evaluating}, applied to a single logit; we apply it per coded bit. The gradient $\nabla_{\boldsymbol{\delta}} \mathcal{L}_{\mathrm{LLR}}$ flows from $\boldsymbol{\ell}_{\mathrm{pre}}$ directly back to $\boldsymbol{\delta}$, bypassing the unrolled BP iterations entirely.

\subsection{Projected Gradient Descent Optimizer} \label{sec:attacks_b} 
We use $\ell_2$-projected gradient descent (PGD) following the framework of Madry \emph{et al.}~\cite{madry2018towards}, with the sign convention reversed for adversarial maximization. The optimization is initialized with a complex Gaussian perturbation projected onto the $\ell_2$-ball boundary of radius $\varepsilon$,
\begin{equation}
\boldsymbol{\delta}^{(0)} = \varepsilon \cdot 
\frac{\boldsymbol{\eta}}{\|\boldsymbol{\eta}\|_2},
\quad \boldsymbol{\eta} \sim \mathcal{CN}(\mathbf{0}, \mathbf{I}_{N_c}),
\label{eq:pgd_init}
\end{equation}
where $\varepsilon > 0$ is the perturbation budget and $N_c$ denotes the codeword length.
At each step $t = 0, 1, \ldots, T-1$, the gradient of the chosen loss function is evaluated at $\boldsymbol{\delta}^{(t)}$, normalized to unit $\ell_2$-norm, scaled by step size $\alpha$, and added to the current iterate. The result is projected onto the $\ell_2$-ball of radius $\varepsilon$,
\begin{equation}
\boldsymbol{\delta}^{(t+1)} = \Pi_{\varepsilon}
\biggl[\boldsymbol{\delta}^{(t)} + \alpha \cdot 
\frac{\nabla_{\boldsymbol{\delta}} \mathcal{L}\bigl(\boldsymbol{\delta}^{(t)}\bigr)}
{\|\nabla_{\boldsymbol{\delta}} \mathcal{L}\bigl(\boldsymbol{\delta}^{(t)}\bigr)\|_2}\biggr],
\label{eq:pgd}
\end{equation}
where the Euclidean projection onto the $\ell_2$-ball of radius $\varepsilon$ is defined as
\begin{equation}
\Pi_{\varepsilon}[\mathbf{u}] = \begin{cases} 
\mathbf{u} & \text{if } \|\mathbf{u}\|_2 \leq \varepsilon, \\ 
\varepsilon \cdot \dfrac{\mathbf{u}}{\|\mathbf{u}\|_2} & 
\text{otherwise}.
\end{cases}
\label{eq:proj}
\end{equation}
The gradient at each PGD step is evaluated on the same channel, noise sample, and codeword used to initialize the per-realization optimization. We do not average over noise or channel during perturbation design; the resulting per-realization perturbations are evaluated either on the matching realization (per-realization white-box attack) or transfer attack; both evaluated in Sec.~\ref{sec:results}.

\subsection{Baselines}\label{sec:attacks_c}

We compare the three types of PGD attacks against two baselines at matched perturbation power. The \textit{Gaussian jammer} samples $\boldsymbol{\delta}_{\mathrm{Gauss}} = \varepsilon \cdot \boldsymbol{\eta}/\|\boldsymbol{\eta}\|_2$, where $\boldsymbol{\eta} \sim \mathcal{CN}(\mathbf{0}, \mathbf{I}_{N_c})$, producing an isotropic perturbation on the $\ell_2$-ball boundary of radius $\varepsilon$. The fast gradient method (FGM) applies one full-budget gradient step on $\mathcal{L}_{\mathrm{E2E}}$ from the same Gaussian initialization as PGD,
\begin{equation}
\boldsymbol{\delta}_{\mathrm{FGM}} = \Pi_{\varepsilon}\!
\biggl[\boldsymbol{\delta}^{(0)} + \alpha_{\mathrm{FGM}} \cdot
\frac{\nabla_{\boldsymbol{\delta}} \mathcal{L}_{\mathrm{E2E}}\bigl(\boldsymbol{\delta}^{(0)}\bigr)}
{\|\nabla_{\boldsymbol{\delta}} \mathcal{L}_{\mathrm{E2E}}\bigl(\boldsymbol{\delta}^{(0)}\bigr)\|_2}\biggr],
\label{eq:fgm}
\end{equation}
where $\boldsymbol{\delta}^{(0)}$ is drawn from~\eqref{eq:pgd_init} and $\alpha_{\mathrm{FGM}} = \varepsilon$ is the full-budget step size. The projection ensures $\|\boldsymbol{\delta}_{\mathrm{FGM}}\|_2 \leq \varepsilon$ when the gradient step exits the $\ell_2$-ball.

\section{Obfuscated Gradients in Unrolled BP }\label{sec:mechanism}
\begin{figure*}[t]
\centering
\includegraphics[width=0.85\textwidth]{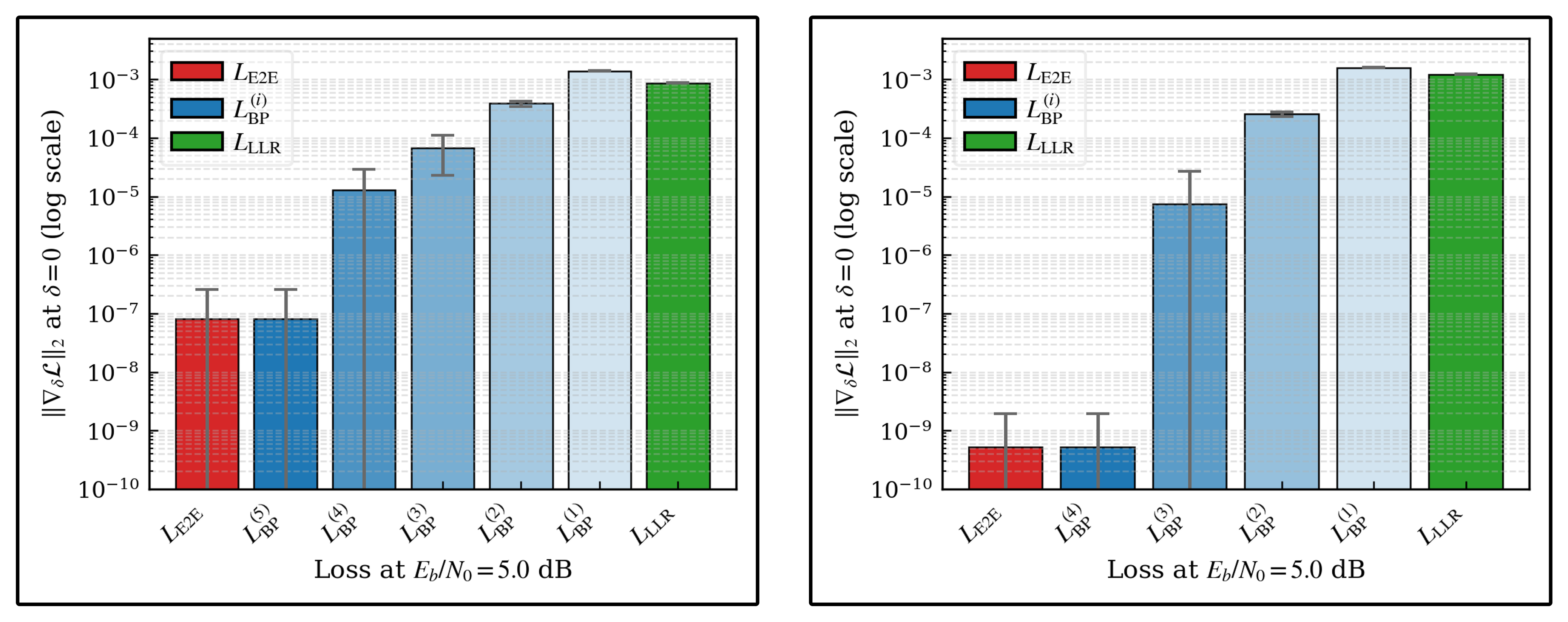}
\caption{Gradient $\ell_2$ norm $\|\nabla_{\boldsymbol{\delta}}\mathcal{L}\|_2$ at $\boldsymbol{\delta}=\mathbf{0}$ for the three loss functions on (left) NeuralReceiver and (right) DeepRx, computed at $E_b/N_0 = 5$\,dB and averaged over 10 random channel realizations (error bars: $\pm 1$ standard deviation).}
\label{fig:gradient_diagnostic}
\end{figure*}

The failure of PGD-E2E to significantly outperform a Gaussian jammer at moderate or low PSR (Sec.~\ref{sec:results}) is consistent with the obfuscated-gradients framework of Athalye \emph{et al.}~\cite{athalye2018obfuscated}. The unrolled BP decoder satisfies the structural condition Athalye \emph{et al.} link to this pathology, namely an iterative computation whose unrolled depth scales with $N_{\mathrm{BP}}$. The mechanism originates in the boxplus check-node update~\cite{hagenauer1996iterative}, an associative operation that combines incoming LLR messages at the check node and is defined for two LLRs as
\begin{equation}
\ell_a \boxplus \ell_b := 2 \tanh^{-1}\!\left( 
\tanh(\ell_a/2) \tanh(\ell_b/2) \right).
\label{eq:boxplus}
\end{equation}
For a check node of degree $d_c$ in the Tanner graph, the outgoing extrinsic LLR is the iterated boxplus of the $d_c - 1$ incoming variable-node messages. We compute the gradient $\ell_2$ norm $\|\nabla_{\boldsymbol{\delta}}\mathcal{L}\|_2$ at $\boldsymbol{\delta}=\mathbf{0}$ for the three loss functions on both architectures, averaged over 10 random channel realizations at $E_b/N_0=5$\,dB. The diagnostic at $\boldsymbol{\delta}=\mathbf{0}$ characterizes the gradient available to an attacker at the operating point of the system, with results shown in Fig.~\ref{fig:gradient_diagnostic}.
%

%
The gradient $\ell_2$ norm of end-to-end loss is at the numerical floor,  $\|\nabla_{\boldsymbol{\delta}}\mathcal{L}_{\mathrm{E2E}}\|_2 \approx 10^{-7}$ for NeuralReceiver and $\approx 10^{-9}$ for DeepRx. By contrast, $\|\nabla_{\boldsymbol{\delta}}\mathcal{L}_{\mathrm{LLR}}\|_2 \approx 10^{-3}$ for both architectures, exceeding $\|\nabla_{\boldsymbol{\delta}}\mathcal{L}_{\mathrm{E2E}}\|_2$ by four and six orders of magnitude for NeuralReceiver and DeepRx, respectively. We localize the masking by computing $\|\nabla_{\boldsymbol{\delta}}\mathcal{L}_{\mathrm{BP}}^{(i)}\|_2 $ for each $i = 1,\ldots, N_{\mathrm{BP}}$. For NeuralReceiver, $\|\nabla_{\boldsymbol{\delta}}\mathcal{L}_{\mathrm{BP}}^{(i)}\|_2$ decreases from $ \approx 10^{-3}$ at $i=1$ to $\approx 10^{-7}$ at $i=5$, where it matches $\|\nabla_{\boldsymbol{\delta}}\mathcal{L}_{\mathrm{E2E}}\|_2 $. For DeepRx, the same convergence is reached one iteration earlier at $i=4$, with $\|\nabla_{\boldsymbol{\delta}}\mathcal{L}_{\mathrm{BP}}^{(4)}\|_2 \approx 10^{-9}$. Each BP iteration attenuates the gradient by approximately one to two orders of magnitude. The convergence of $\|\nabla_{\boldsymbol{\delta}}\mathcal{L}_{\mathrm{BP}}^{(N_{\mathrm{BP}})}\|_2 $ and $\|\nabla_{\boldsymbol{\delta}}\mathcal{L}_{\mathrm{E2E}}\|_2 $ at the same floor is the central piece of evidence. The gradients of both loss functions traverse the same chain of $N_{\mathrm{BP}}$ saturating boxplus check-node updates. The gradient masking is therefore attributable to the BP chain rather than to the receiver, consistent with the structural prediction of the obfuscated-gradients framework.

Because $\mathcal{L}_{\mathrm{BP}}$ sums the per-BP-iteration loss function terms without normalization, $\nabla_{\boldsymbol{\delta}}\mathcal{L}_{\mathrm{BP}}$ is dominated by the early-BP-iteration term regardless of how strongly later iterations are masked. PGD therefore operates on a gradient with negligible contribution from later iterations. The gradient of the pre-decoder margin loss $\mathcal{L}_{\mathrm{LLR}}$ bypasses the BP graph entirely. The gradient flows from the pre-decoder LLR vector $\boldsymbol{\ell}_{\mathrm{pre}}$ directly back to $\boldsymbol{\delta}$, with no boxplus check-node updates along the path. The cross-architecture invariance of the gradient diagnostic is itself a finding. Despite their architectural differences, NeuralReceiver and DeepRx exhibit the same gradient masking pattern and the same per-iteration gradient decay rate. Gradient masking in this setting is therefore a property of the BP decoding, independent of the receiver architecture.

\section{Results}\label{sec:results}
We configure the system model of Sec.~\ref{sec:system} with the following parameters. The carrier frequency is $f_c = 3.5$\,GHz. The receiver is equipped with $N_r = 2$ antennas. The OFDM resource grid spans $T = 14$ OFDM symbols and $F = 128$ subcarriers, yielding total complex dimensionality $N_{\mathrm{c}} = 3584$. The CDL-C channel is configured with 100\,ns RMS delay spread. The 5G NR LDPC code has $k = 1392$ information bits and codeword length $n = 2784$, corresponding to code rate $R = 1/2$. As for the attack parameters, we set $m=2.0$, $T_{\mathrm{PGD}} = 100$ steps, and $\alpha = \varepsilon/10$. 
We configure the perturbation batch size to be $512$. 
All experiments are performed at $E_b/N_0 = 5$\,dB, with each BLER point evaluated until 200 block errors or $2.048\times 10^6$ blocks. Large pre-decoder LLRs at $E_b/N_0 = 5$\,dB drive the BP decoder into deep saturation, maximizing gradient masking. The attack efficacy demonstrated under these conditions necessarily extends to lower SNR regimes.

We isolate the channel-transfer behavior of each attack by comparing two evaluation modes. Case~A (same-channel) evaluates each per-realization perturbation $\boldsymbol{\delta}_i^{\ast}$ on the channel realization $\mathbf{H}_i$ on which it was optimized. Case~B (transfer) applies the designed perturbation set unchanged to a disjoint batch $\{\mathbf{H}'_j\}$, where PGD is executed once per PSR point. The gap between Case~A and Case~B BLER measures the extent to which 
the attack overfits to the channel over which it is designed. A channel-invariant adversarial direction yields comparable BLER under both modes. Case~B represents the operationally relevant scenario for an adversary who commits to a perturbation prior to observing the channel realization.

\begin{figure*}[t]
\centering
\includegraphics[width=0.85\textwidth]{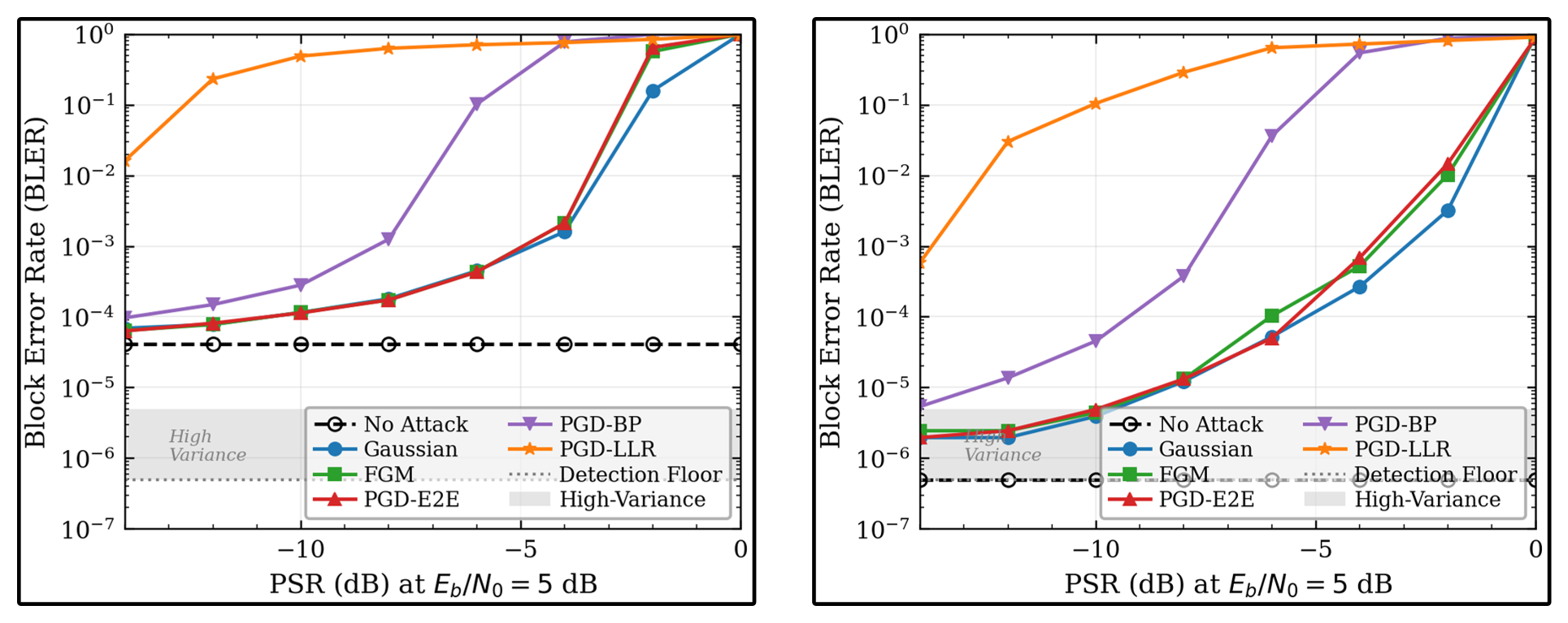} \caption{Block error rate (BLER) as a function of PSR under $\ell_2$-constrained adversarial perturbations for (left) NeuralReceiver and (right) DeepRx, evaluated in Case B (transfer mode). Operating point is $E_b/N_0 = 5$\,dB with 5G NR LDPC ($k=1392$, $n=2784$, rate $1/2$), CDL-C channel at $100$\,ns RMS delay spread, and QPSK modulation.}
\label{fig:bler_vs_psr}
\end{figure*}

Fig.~\ref{fig:bler_vs_psr} shows Case~B BLER versus PSR for both architectures; Table~\ref{tab:transferability} provides the Case~A--Case~B comparison. PGD-LLR outperforms PGD-BP, PGD-E2E, and the Gaussian jammer across the entire PSR range, reaching the same BLER as PGD-BP at $4$--$8$\,dB lower PSR. PGD-E2E, FGM, and the Gaussian jammer trace nearly identical BLER curves, all remaining at the no-attack floor for PSR $\leq -8$\,dB and rising together to near-unity BLER at PSR $= 0$\,dB. PGD-E2E (100 iterative steps) and FGM (1 step) span the full range of optimization effort yet yield indistinguishable performance, which is a diagnostic signature of gradient masking in~\cite{athalye2018obfuscated}.

The ordering PGD-E2E~$\ll$~PGD-BP~$\ll$~PGD-LLR holds for both architectures despite the $7.6 \times$ difference in the number of parameters. On NeuralReceiver, PGD-LLR reaches BLER $\approx 0.5$ at PSR $= -10$\,dB, where PGD-BP requires PSR $= -5$\,dB and the Gaussian jammer requires PSR $= -1.5$\,dB. On DeepRx, the corresponding crossings are shifted by $1$--$3$\,dB toward higher PSR, with the shift larger for the more effective attacks. PGD-LLR exceeds the Gaussian jammer by $10^2$--$10^4\times$ in BLER across the moderate PSR range, with the largest gap of $\approx 10^4\times$ on DeepRx at PSR $= -8$\,dB. PGD-BP exceeds the Gaussian jammer by approximately $10\times$ at PSR $= -8$\,dB on NeuralReceiver and $30\times$ on DeepRx, growing to $> 100\times$ at PSR $= -6$ to $-4$\,dB.

\begin{table}[t]
\centering
\caption{$\mathrm{Case\;B\;(transfer)\;BLER\;for\;PGD-BP\;and\; PGD-LLR\;at\;E_b/N_0}$ 
$\mathrm{=5dB.\;Case\;A\;BLER\;is\;}$ $\approx \mathrm{1.0\;for\;both\; \;attacks\;at\;all\;PSR\;values.}$}
\label{tab:transferability}
\renewcommand{\arraystretch}{1.15}
\setlength{\tabcolsep}{4pt}
\small
\begin{tabular}{ll cccc}
\toprule
Receiver & Attack & \multicolumn{4}{c}{PSR (dB)} \\
\cmidrule(lr){3-6}
& & $-12$ & $-8$ & $-4$ & $0$ \\
\midrule
\multirow{2}{*}{NRx} 
& PGD-BP  & $1.4\!\times\!10^{-4}$ & $1.2\!\times\!10^{-3}$ & $0.77$ & $1.0$ \\
& PGD-LLR & $0.23$ & $0.63$ & $0.76$ & $0.98$ \\
\midrule
\multirow{2}{*}{DeepRx} 
& PGD-BP  & $1.37\!\times\!10^{-5}$ & $3.7\!\times\!10^{-4}$ & $0.54$ & $0.97$ \\
& PGD-LLR & $3.0\!\times\!10^{-2}$ & $0.29$ & $0.72$ & $0.90$ \\
\bottomrule
\end{tabular}
\end{table}

PGD-BP and PGD-LLR both achieve Case A BLER $\approx 1.0$ across the PSR sweep, confirming that both attacks are highly effective on the channels for which they are optimized. PGD-BP Case B BLER falls by up to five orders of magnitude at low PSR, with $\approx 10^{-4}$ for NeuralReceiver and $\approx 10^{-5}$ for DeepRx at PSR $= -12$\,dB. PGD-BP overfits to the crafting channel, producing a channel-conditioned adversarial direction. PGD-LLR retains a substantial fraction of its Case A BLER in Case B at moderate PSR, reaching $0.63$ for NeuralReceiver and $0.29$ for DeepRx at PSR $= -8$\,dB. The PGD-LLR Case B BLER exceeds that of PGD-BP by two to three orders of magnitude at PSR $\in \{-12, -8\}$\,dB. Although the adversary has no knowledge of the resource-grid layout, the $\mathcal{L}_{\mathrm{LLR}}$-optimized perturbation concentrates on the OFDM symbols carrying pilots, which is consistent with \cite{ulbricht2024pyjama}. Since these positions are fixed independently of the channel realization, the resulting direction is channel-invariant by construction. At very low PSR values, the perturbation budget is highly constrained, and both attacks transfer poorly. At high PSR ($\geq -4$\,dB), both attacks reach near-unity Case~B BLER and the performance gap narrows. The moderate-PSR regime is therefore the critical region in which the available budget separates channel-invariant from channel-conditioned perturbation directions.
\section{Conclusion}\label{sec:conclusion}
We have shown that standard $\ell_2$-FGM and PGD attacks based on $\mathcal{L}_{\mathrm{E2E}}$ fail to outperform a matched-power Gaussian jammer against NNRs with unrolled 5G NR LDPC decoders. The failure is the diagnostic signature of obfuscated gradients~\cite{athalye2018obfuscated}. The gradient norm decays by more than four orders of magnitude across the ten unrolled BP iterations. Two structurally independent losses restore attack efficacy, increasing BLER by one to four orders of magnitude over PGD-E2E. $\mathcal{L}_{\mathrm{BP}}$ routes the gradient through early unsaturated BP iterations, while the gradient of $\mathcal{L}_{\mathrm{LLR}}$ bypasses the BP graph entirely. Perturbations crafted with $\mathcal{L}_{\mathrm{LLR}}$ transfer across channel realizations with up to three orders of magnitude higher BLER than those crafted with $\mathcal{L}_{\mathrm{BP}}$ at low PSR. All findings replicate for both NeuralReceiver and DeepRx.

Several extensions remain open. The threat model can be extended to route the perturbation through a physical adversary channel rather than adding it directly at the receiver antennas. 
\section*{Acknowledgment}
This work is Co-funded by the European Union (EU) under Grant Agreement
101191936. Views and opinions expressed are, however, those of the authors only and do not necessarily reflect those of all SUSTAIN-6G consortium parties nor those of the EU or the SNS JU.
Neither the EU nor the granting authority can be held responsible for them.

\bibliographystyle{ieeetr}
\bibliography{Bibliography}

\end{document}